\documentclass[useAMS,usenatbib]{mn2e}

\usepackage {graphicx}
\usepackage {times}
\def\aj{\,{AJ}}
\def\apj{\,{\rm ApJ}}
\def\apjl{\,{\rm ApJL}}
\def\apjs{\,{\rm ApJS}}

\def\pasj{\,{\rm PASJ}}

\def\aap{\,{\rm A\&A}}

\def\mnras{\,{\rm MNRAS}}
\def\kms{\, km s$^{-1}$ }

\title[Velocity Dispersion of 335 Galaxy Clusters]
{Velocity Dispersion of 335 Galaxy Clusters Selected from the Sloan Digital Sky Survey: Statistical Evidence for Dynamical Interaction, and Against Ram-Pressure Stripping}
\author[T. Goto]
{Tomotsugu Goto$^{1}$\thanks{E-mail:tomo@jhu.edu}
  \\
  $^{1}$ Department of Physics and Astronomy, The Johns Hopkins
  University, 3400 North Charles Street, Baltimore, MD 21218-2686, USA
}

\begin{document}

\pagerange{\pageref{firstpage}--\pageref{lastpage}} \pubyear{2004}

\maketitle

\label{firstpage}

\begin{abstract}

 There has been plenty of observational evidence of cluster galaxy evolution such as the Butcher-Oemler effect and the morphology-density relation. However, it has been difficult to identify the origin of the cluster galaxy evolution. It has been simply difficult to trace the complicated process of galaxy evolution with several giga years of timescale, using the observation which only brings us with information of a single epoch. Here we show that gravitational interaction/friction between galaxies is the statistically dominant physical mechanism responsible for the cluster galaxy evolution, and that the well-favored ram-pressure stripping by the cluster gas is not statistically driving the cluster galaxy evolution. 
   We have constructed the largest composite cluster with 14548 member galaxies out of 335 clusters with $\sigma>300$ km s$^{-1}$ carefully selected from the Sloan Digital Sky Survey. By measuring velocity dispersions of various subsamples of galaxies in this composite cluster,  we found that bright cluster galaxies ($Mz<-23$) have significantly smaller velocity dispersion than faint galaxies ($Mz\geq-23$), with much greater precision than previous results (e.g., Adami et al.). We interpret this as direct evidence of the dynamical interaction/friction between cluster galaxies, where massive galaxies lose their velocity through the energy equipartition during the dynamical interaction/friction with less massive galaxies. We also found that star-forming late-type galaxies have a larger velocity dispersion than passive late-type galaxies. This is inconsistent with the ram-pressure stripping model; since the ram-pressure is proportional to $\sigma v^2$ (i.e., stronger for galaxies with high velocity), the ram-pressure stripping cannot explain the observed trends of passive (evolved) galaxies having low velocity rather than high velocity. On the other hand, the result is again consistent with the dynamical galaxy-galaxy interaction/friction, where more evolved (passive) galaxies lose their velocity through dynamical interaction/friction.

\end{abstract}


\section{Introduction}\label{intro}


 It has been well-known that cluster galaxies evolve with redshift. 
 Butcher \& Oemler (1978,1984) found that fractions of blue galaxies in clusters increase  with increasing redshift (see also Rakos, Schombert 1995; Couch et al. 1994,1998;  Margoniner, de Carvalho 2000; Margoniner et al. 2001; Ellingson  et  al. 2001; Kodama \& Bower 2001; Goto et al. 2003a,2004a; but also see Andreon et al. 1999,2004).
      In a modern version, high redshift clusters 
  are known to  have larger fractions of star-forming galaxies than local clusters  (Postman, Lubin, \& Oke 1998,2001; Finn, Zaritsky, \& McCarthy 2004).

      It is also known that cluster galaxies show the morphological evolution as well, i.e., fractions of S0 galaxies are found to be lower in high redshift clusters (Dressler et al. 1997; van Dokkum et al. 2000; Fasano et al. 2000; Jones, Smail, \& Couch 2000;  Fabricant et al. 2000;  Goto et al. 2003b;  also see Andreon et al. 1998).
  Goto et al. (2003a,2004a) used 516  clusters found in the Sloan Digital Sky Survey (Goto et al. 2002a,b) to verify this effect with greater statistical significance.


    These numerous observational evidences suggest that a certain
   physical mechanism(s) is changing the morphology and star formation rate (SFR) of cluster
   galaxies as a function of redshift. 
   However to date, it has been difficult to specify what physical mechanisms determine
    morphology and SFR of galaxies in clusters.
 Since observations only provide us with a snapshot of the cosmic history, it has been difficult to trace the complicated process of galaxy evolution using traditional observational information on photometry and spectral energy distribution.
 It is also worth noting that E+A (k+a or post-starburst) galaxies,
which often thought to be cluster-related, are found to have their origin in merger/interaction with accompanying galaxies (Goto et al. 2003d,e,2004b,c), 
 and thus E+A galaxies are not likely to be
a product of the morphological transition in cluster regions.

Various physical mechanisms have been proposed to explain the cluster galaxy evolution. One of the most popular candidates is the ram-pressure stripping
 of cold gas by the intra-cluster medium (Gunn \& Gott 1972; Farouki \& Shapiro 1980; Kent 1981; Fujita \& Nagashima 1999;  Abadi, Moore \& Bower 1999; Quilis, Moore \& Bower 2000; Fujita 2004; Fujita \& Goto 2004). In this ram-pressure stripping model, a galaxy infalling from outside of a cluster suffers from the ram-pressure by the intra-cluster medium (ICM), and then, the cold gas in the disk of the galaxy becomes stripped away. Once the cold gas is stripped, the star formation, especially that in the disk where most of the cold gas exists, rapidly ceases in the time scale of $\sim50$Myr (Quilis et al. 2000). As a result, the galaxy colour becomes red and its morphology becomes similar to a S0 galaxy (e.g., Fujita \& Nagashima 1999). 
 
 A similarly favored model is the hot halo gas stripping (strangulation, suffocation or starvation). In this model, hot gas resides in a halo of a galaxy supplying the cold gas to the disk of the galaxy. The hot gas can be quickly stripped by the ram-pressure from the ICM even with the smaller pressure that cannot remove the more strongly-bound cold gas in the disk. This is because the hot gas resides in the outer regions (halo) of the galaxy and the gas density is much smaller. Once the hot gas is stripped, the supply to the cold disk gas stops. As a result, the cold disk gas is gradually consumed by the star formation. Since the cold gas is the ingredient of stars, the star formation activity of the galaxy also gradually declines. Therefore, the timescale of the decline of the star formation is longer than that of the cold gas stripping (Larson et al. 1980).  This hot gas stripping model has been favored over the cold gas stripping model since recently red cluster galaxies have been
 found beyond the virial radius of the clusters where the ICM density is too low for the cold gas stripping to happen (Kodama et al. 2001b; Goto et al. 2003c; Treu et al. 2003; Tanaka et al. 2004; Goto et al. 2004d).


 Galaxy velocity dispersion in clusters can bring a new observational constraint on the subject since it can directly probe the kinematic state of cluster galaxies. 
  Previously, Zabludoff \& Franx (1993) reported the velocity distribution of spiral galaxies is different from those of other morphological populations in 3/6 clusters they observed.  It is later reported that spiral galaxies have a greater velocity dispersion than ellipticals in nearby clusters (Sodre et al. 1989; Stein 1997; Adami et al. 1998,2000). In addition, Adami et al. (1998) reported that galaxies of different types have different velocity dispersion profiles, being steeper for late type galaxies. 
Carlberg et al. (1997) reported that red galaxies have a smaller rms velocity dispersion by a factor of 1.31$\pm$0.13. 
 Biviano et al. (1997) found that velocity dispersion of emission line galaxies is, on average, 20\% larger than that of other galaxies (see also Biviano et al. 2002; Lares et al. 2004). 
 However, in order to investigate subtle difference in kinematics of different subsamples of cluster galaxies in greater accuracy, large number of spectra are required, especially, when measuring a velocity dispersion of rare population of galaxies (e.g., red late-types in Goto et al. 2003c; Yamauchi \& Goto 2004). In the literature, there exists a handful of clusters with $\sim$100 of spectroscopic members (e.g., Czoske et al. 2002; Gal \& Lubin 2004), however, even with these extreme cases, there were not enough number of galaxies to measure velocity dispersion of subsamples of cluster galaxies precisely.
 In order to overcome this problem, we have created a catalog of 335 galaxy clusters with $\sigma>300$ km s$^{-1}$ using the 250,000 spectra of Sloan Digital Sky Survey (SDSS) Data Release 2 (Abazajian et al. 2004). Due to the large area coverage of the SDSS,  a composite cluster created with this cluster catalog has an extremely large number of 14548 member galaxies. And thus, it is an ideal sample to measure velocity dispersion of different subsamples of galaxies. The statistics based on this sample is unparalleled to any of the previous work. In addition to high precision measurements of previously suggested subsamples such as separated by morphology, emission lines, or magnitude, we measure velocity dispersion of subsamples separated by both morphology and emission lines, which will bring new insights as described later.  Before the SDSS, it has been simply impossible to measure velocity dispersion of subsamples using only dozens of galaxy clusters.

 This paper is organized as follows: In Section \ref{data}, we describe the SDSS data and construct a catalog of 335 galaxy clusters; In Section \ref{analysis}, we measure the velocity dispersion of various sub-samples of cluster members to investigate mass dependence (Section \ref{absolute_magnitude}), morphology dependence (Section \ref{morpholgy_result}), colour dependence (Section \ref{color_result}), the dependence on both morphology and colour (Section \ref{morphology_color}), and SFR dependence (Section \ref{morphology_SFR}). We find evidence that massive galaxies have smaller velocity dispersion than less massive galaxies, and that passive late-type galaxies have smaller velocity dispersion than star-forming late-type galaxies; In Section \ref{systematics}, we discuss possible systematics effects. In Section \ref{physical}, we discuss the physical implications of our results on the physical mechanisms that govern cluster galaxy evolution; In Section \ref{conclusion}, we summarize our work and findings.
 Unless otherwise stated, we adopt the best-fit WMAP cosmology:
 $(h,\Omega_m,\Omega_L) = (0.71,0.27,0.73)$
 (Bennett et al. 2003).

\section{Data}\label{data}

\subsection{The SDSS DR2 Data}

 We have downloaded the SDSS DR2 galaxy catalog using the SDSS SQL tool (Abazajian et al. 2004). We only use those objects classified as galaxies ({\tt type=3}, i.e., extended) with spectroscopically measured redshift of $z>0.03$. These two criteria can almost entirely eliminate contamination from mis-classified stars and HII regions in nearby galaxies. The resulting number of galaxies used is 253,497. 

\subsection{Constructing a Cluster Catalog}
For each galaxy in this sample, we measure the number of neighbor galaxies within 1.5 Mpc in angular direction and $\pm1000$ \kms in the line-of-sight direction. If a galaxy has a neighbor, neighbors of the neighbor are searched using the same linking length of 1.5 Mpc and $\pm1000$ \kms until no further neighbors are found. It is estimated that we require at least 20 galaxies to securely measure velocity dispersion of an individual cluster (e.g., Biviano et al. 1997). Therefore, only when the group has more than 20 linked neighbors, we regard it as a cluster.  

Once a cluster is found, we measure the redshift and velocity dispersion of the cluster using the biweight estimator with 3 $\sigma$ clipping since it is statistically more robust and efficient than the standard mean in computing the central location of a data-set (Beers et al. 1990). If the cluster has a velocity dispersion greater than 300 \kms, we include 3 $\sigma$ clipped members into a composite cluster. The relative velocity of each member galaxy to the cluster redshift is normalized by the velocity dispersion of the cluster to be included into the composite cluster. 
 In this process, we have removed galaxies with a clustocentric radius of more than 2 virial radius in angular direction using virial radii computed in Goto (2005b). This process nicely limits analysis to the regions with similar gas density when measuring velocity dispersion of various subsamples of galaxies in the composite cluster.

 Among the SDSS DR2 sample, we have found 335 clusters with $>20$ members and $\sigma>300$ \kms. This cluster catalog contains one of the largest number of clusters with measured velocity dispersion.  Among the member galaxies, we adopted the position of the brightest (in $z$-band) cluster galaxy as an angular position of the cluster. This cluster catalog contains positions (RA,DEC,J2000), redshift, number of member galaxies, and measured velocity dispersion of the clusters, and is presented electronically \footnote{http://acs.pha.jhu.edu/$\sim$tomo}. The created composite cluster has 14548 member galaxies,  which is by two orders of magnitude larger than a typical number of member galaxies available in the literature. Hereby, we have obtained the first opportunity to investigate velocity dispersion of various subsample of cluster galaxies with unprecedented high precision.
We have re-measured velocity dispersion of this cluster obtaining $\sigma_{normalized}=$ 0.9937$^{+0.0058}_{-0.0057}$. 
We take a conservative approach in computing errors using Danese, de Zotti, \& di Tullio(1980), which usually returns larger values than a jack knife estimator. 
Nonetheless, the errors are extremely small due to the large number of member galaxies, making the composite cluster an ideal sample to investigate velocity dispersion of different subsamples of galaxies.

\section{Analysis \& Results}\label{analysis}

 In the following subsections,  we measure velocity dispersion of different sub-samples of cluster galaxies using the biweight estimator and the errors of Danese et al. (1980). In subsection \ref{absolute_magnitude}, we investigate  velocity dispersion as a function of absolute magnitude. In subsections \ref{morpholgy_result}-\ref{morphology_SFR}, we separate subsamples using morphology, colour and SFR.

\subsection{Velocity Dispersion as a Function of Absolute Magnitude}\label{absolute_magnitude}

\begin{figure}
\includegraphics[scale=0.65]{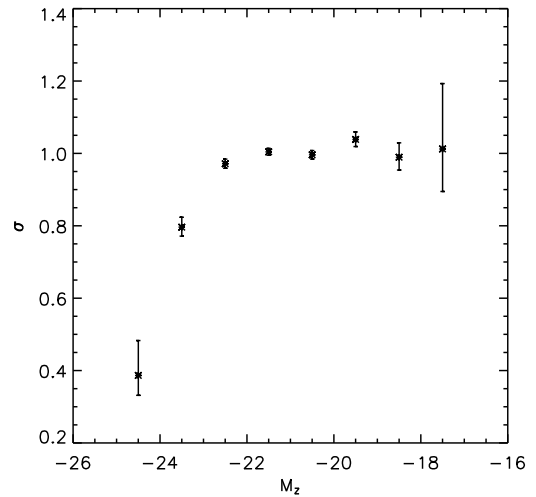}
\caption{
\label{fig:absolute_mag} 
 Velocity dispersion of the composite cluster as a function of absolute magnitude in $z$-band. Dispersions are computed using galaxies in $\pm0.5$ mag bins. The result is consistent with Adami et al. (1998), but presented with much greater precision. 
}
\end{figure}

 In this subsection, we investigate velocity dispersion as a function of $z$-band absolute magnitude ($M_z$). 
 We use the $z$-band absolute magnitude as the closest approximation to the galaxy mass among the 5 SDSS colour bands.  It is estimated the variation in the mass-to-light ratio is only a factor of $\sim 3$ in $z$-band (Kauffmann et al. 2003).
 In the following, we use absolute magnitude in $z$-band as a galaxy mass estimator.  We use Petrosian magnitude corrected for galactic extinction using a reddening map of Schlegel, Finkbeiner \& Davis (1998). We use $k$-correction given in Blanton et al. (2003;v3\_2)  to calculate absolute magnitudes.  
 Fig. \ref{fig:absolute_mag} shows the measured velocity dispersion as a function of $M_z$. An immediately notable feature is that at $M_z<-23.0$, the velocity dispersion decrease toward brighter absolute magnitude. 
 Also, it is seen that between $M_z=-20.0$ and $-$23.0, the velocity dispersion are consistent with a constant. When we divide the sample using morphology and colour in the following sections, we limit galaxies to $-22.5\leq M_z \leq -20.5$ in order to minimize the mass (or $M_z$) dependent bias.

\subsection{Velocity Dispersion as a Function of Galaxy Morphology}\label{morpholgy_result}

 Next, we investigate velocity dispersion for different morphological types of galaxies. We use the ratio  of Petrosian 50\% light radius to Petrosian 90\% light radius, $Cin$, measured in $r$-band image to quantify galaxy morphology.  The analysis will reveal if kinematic structures of galaxies depend on morphological galaxy types. To avoid magnitude dependent bias, we only use galaxies with $-22.5\leq M_z \leq -20.5$, where velocity dispersion is constant  as a function of $M_z$ (Fig. \ref{fig:absolute_mag}). 
We regard galaxies with $Cin>0.4$ as late-type galaxies and $Cin\leq 0.4$ as early-type galaxies. This parameter $Cin$ is known to be well correlated with eye-classified morphology (Shimasaku et al. 2001; Strateva et al. 2001), and thus has been used in many previous studies (e.g., Gomez et al. 2003; Kauffmann et al. 2004). 

 We show the results in Table \ref{tab:velocity_dispersion_composite}, where the late-type galaxies ($Cin> 0.4$) has $\sigma=$1.040$_{-0.012}^{+0.012}$ and the early-type galaxies ($Cin\leq 0.4$) has $\sigma=$0.963$_{-0.007}^{+0.007}$. Since both samples have more than 3000 galaxies, the difference is quite significant ($>4\sigma$ significance level). The difference is larger than the magnitude dependence we saw in the $-22.5\leq M_z \leq -20.5$ mag range in Fig. \ref{fig:absolute_mag}. Therefore, the results indicate that late-type galaxies have larger velocity dispersion than early-type galaxies.  We discuss physical implications of the result in Section \ref{discussion}.

\begin{table*}
\caption{
Normalized Velocity dispersion ($\sigma$) for subsamples of galaxies in the composite cluster. The number of spectroscopic galaxies in each sub-sample is listed as $N_{galaxies}$. The composite cluster consists of large number of 14548 members in 335 clusters. 
}\label{tab:velocity_dispersion_composite}
\begin{center}
\begin{tabular}{lrr}
\hline
Sample &  $\sigma$ (normalized)  & $N_{galaxies}$   \\
\hline
\hline
 All    & 0.9937$^{+0.0058}_{-0.0057}$ & 14548\\
\hline
 Early-Type ($Cin\leq 0.4$)      & 0.965$_{-0.007}^{+0.007}$   &  7987 \\
 Late-Type   ($Cin> 0.4$)        & 1.040$_{-0.012}^{+0.012}$   &  3396 \\
\hline
 Red ($u-r\geq 2.22$)       & 0.961$_{-0.007}^{+0.007}$    &  9003 \\
 Blue  ($u-r<2.22$)         & 1.085$_{-0.015}^{+0.016}$    &  2380 \\
\hline
 Passive  ($SFR<2 M_{\odot} yr^{-1}$)                & 0.960$_{-0.007}^{+0.007}$    &  8183 \\
 Star-Forming ($SFR\geq 2 M_{\odot} yr^{-1} $)       & 1.128$_{-0.047}^{+0.054}$    &  246 \\
\hline
 Red Late-Type  ($u-r\geq 2.22$ \& $Cin> 0.4$)     &  0.983$_{-0.017}^{+0.018}$    & 1528  \\
 Blue Late-Type ($u-r< 2.22$ \& $Cin> 0.4$)    & 1.084$_{-0.017}^{+0.018}$    & 1868   \\
\hline
 Passive Late-Type      ($SFR<2 M_{\odot} yr^{-1}$ \& $Cin> 0.4$)     &  1.010$_{-0.019}^{+0.020}$    & 1329  \\
 Star-Forming Late-Type ($SFR\geq 2 M_{\odot} yr^{-1} $ \& $Cin> 0.4$)    & 1.248$_{-0.075}^{+ 0.092}$    & 114 \\
\hline
 Red Early-Type  ($u-r\geq 2.22$ \& $Cin\leq 0.4$)    &  0.956$_{-0.007}^{+0.007}$ & 7475\\
 Blue Early-Type ($u-r< 2.22$ \& $Cin\leq 0.4$)  &  1.092$_{-0.032}^{+0.035}$  & 512\\
\hline
 Passive Early-Type  ($SFR<2 M_{\odot} yr^{-1}$ \& $Cin\leq 0.4$)    &  0.950$_{-0.008}^{+0.008}$ & 6854\\
 Star-Forming Early-Type ($SFR\geq 2 M_{\odot} yr^{-1}$ \& $Cin\leq 0.4$)  &  1.009$_{-0.057}^{+0.068}$  & 132\\
\hline
\end{tabular}
\end{center}
\end{table*}

%

\subsection{Velocity Dispersion as a Function of Galaxy Colour}\label{color_result}

 Next, we investigate velocity dispersion for populations of galaxies with different colour. The rest-frame colour of galaxies generally correlates well with the star formation activity of galaxies (e.g., Kennicutt 1992). Therefore, by investigating velocity dispersion as a function of galaxy colour, we can test if the dynamical state of each cluster galaxy affects the star-formation activity of individual galaxies. 
 We intend to separate blue and red galaxies using the restframe $u-r=2.22$. Strateva et al. (2001) showed that this colour separates early- and late-type galaxies well at $z<0.4$. We use the same magnitude range of $-22.5\leq M_z \leq -20.5$ to minimize the mass (or $M_z$) dependent bias.

The resulting velocity dispersion of blue/red galaxies in the composite cluster are presented in Table \ref{tab:velocity_dispersion_composite}. The blue galaxies in the composite cluster has a velocity dispersion of $\sigma=$1.085$_{-0.015}^{+0.016}$, whereas the red galaxies has $\sigma=$0.961$_{-0.007}^{+0.007}$. Therefore, the velocity dispersion is larger for the blue galaxies  with greater than 5 $\sigma$ significance.  

However, since  galaxy morphology and galaxy colour is well correlated in a sense that earlier type galaxies have redder colour, it is difficult to determine whether the velocity dispersion is more sensitive to galaxy colour or morphology.
 It is essential to use morphology and colour at the same time to answer the question.

\subsection{Velocity Dispersion as a Function of Galaxy Colour and Morphology}\label{morphology_color}

 In this section, we use colour and morphology at the same time in order to distinguish which one of the two is better correlated with the velocity dispersion. 
Since it has been suggested that the timescale of the morphological evolution and spectral (colour) evolution is considerably different in galaxy evolution (Poggianti et al. 1999; Goto et al. 2003c, 2004a; McIntosh, Rix, \& Caldwell 2004), this investigation may be able to bring insight on the origin of the cluster galaxy evolution. 

  In Table \ref{tab:velocity_dispersion_composite}, we show velocity dispersion of red late-type galaxies and blue late-type galaxies in the composite cluster. In the same absolute magnitude range, we select red late-type as galaxies with $u-r>2.22$ and $Cin>0.4$, and blue late-type as galaxies with  $u-r\leq 2.22$ and $Cin>0.4$. 
In Table \ref{tab:velocity_dispersion_composite}, the red late-type galaxies ($u-r\geq 2.22$ \& $Cin> 0.4$)  have $\sigma= 0.983^{+0.018}_{-0.017}$km s$^{-1}$ and the blue late-type galaxies ($u-r< 2.22$ \& $Cin> 0.4$) have $\sigma=1.084^{+0.018}_{-0.07}$km s$^{-1}$,  i.e., the blue late-type galaxies have larger velocity dispersion at a 2 $\sigma$ significance level.   A Kolomogorov-Smirnov test also shows that these two distributions are different with greater than 96\% significance level. 

 Next, we examine the difference in velocity dispersion between red early-type galaxies  ($u-r\geq 2.22$ \& $Cin\leq 0.4$) and blue early-type galaxies ($u-r< 2.22$ \& $Cin\leq 0.4$). As is shown in Table \ref{tab:velocity_dispersion_composite}, we found that red early-type galaxies have larger velocity dispersion ($\sigma=0.956_{-0.007}^{+0.007}$) than blue ellipticals ($\sigma=1.092_{-0.032}^{+0.035}$) at a $3\sigma$ significance level. 

 Since the same morphological type of galaxies with different colours have different velocity dispersion, it is suggested that velocity dispersion may be better correlated with galaxy colour than galaxy morphology.

\subsection{Velocity Dispersion as a Function of Galaxy SFR and Morphology}\label{morphology_SFR}

 In this subsection, we repeat analyses in Sections \ref{color_result} and \ref{morphology_color} using galaxy SFR instead of colour. This analysis brings important implication on the ram-pressure stripping model since the time scale in which H$\alpha$ line disappear ($\sim$50 Myr) after the end of the star-formation is much shorter than that of colour change ($\sim$1 Gyr). The SFR is measured from the H$\alpha$ emission lines corrected for dust extinction, stellar absorption and the aperture size. AGNs are removed from this analysis using Fig.1 of Goto (2005b). The detailed description in measuring SFR is presented in Goto (2005b). We separate star-forming and passive galaxies at SFR of 2 $M_{\odot} yr^{-1}$. We have checked that a slight change in this criterion does not change the results. As is shown in Table \ref{tab:velocity_dispersion_composite}, star-forming galaxies have larger velocity dispersion ($\sigma=1.128_{-0.047}^{+0.054}$) than passive galaxies ($\sigma=0.0960_{-0.007}^{+0.007}$). This result is consistent with that in Section \ref{color_result}, where the velocity dispersion was larger for blue galaxies than red galaxies. Next, we measure velocity dispersion of star-forming late-type galaxies ($SFR \geq 2 M_{\odot} yr^{-1}$ \& $Cin> 0.4$) and passive late-type galaxies($SFR<2 M_{\odot} yr^{-1}$ \& $Cin> 0.4$). Being consistent with the result in Section \ref{morphology_color}, star-forming late-type galaxies have larger velocity dispersion  ($\sigma=1.248_{-0.075}^{+ 0.092}$) than passive late-type galaxies ($\sigma=1.010_{-0.019}^{+0.020}$). Similarly,   star-forming early-type galaxies have larger velocity dispersion  ($\sigma=1.009_{-0.057}^{+0.068}$) than passive early-type galaxies ($\sigma=0.950_{-0.008}^{+0.008}$).

\section{Discussion}\label{discussion}

\subsection{Possible Systematic Effects}\label{systematics}

  In Section \ref{analysis}, we found that massive galaxies have smaller velocity dispersion than less massive galaxies, and that blue late-type galaxies have larger velocity dispersion than the red late-type galaxies in the composite cluster.
We first consider possible systematic effects that might mimic the observed results.

In the cluster regions, it is well-known that red/elliptical galaxies are dominant in the cluster cores and the blue/spiral galaxies are more numerous in the outskirts (the morphology-density relation; Postman et al. 1984; Goto et al. 2003b). Therefore, it is a concern that the difference in the spatial distributions between our samples might cause the observed trend in the velocity dispersion.
 However, in the cluster outskirts, due to the increase in the distance in the angular direction from the core, galaxies lose the velocity in the line-of-sight direction, reducing the observed velocity dispersion. Therefore, observing more blue galaxies in the outskirts only weakens the observed trend of blue galaxies having larger velocity dispersion. 

Due to the physical size of the fiber spectrograph of the SDSS, the adjacent fibers cannot be closer than 55 arcsec on the sky. Consequently, if there are more than one galaxy with $r<17.77$ within 55 arcsec on the sky, only one of them is spectroscopically observed by the SDSS and the rest are missed from the sample. Although the SDSS tries to compensate for this fiber collision problem in the area where more than one spectroscopic plates are assigned, unfortunately, this effect is more severe in the dense regions such as cluster cores. However, due to the uniform survey criteria of the SDSS ($r<17.77$), the missing galaxies are randomly selected, and thus, unbiased against colour or morphology.  Therefore, the fiber collision problem only reduces the number of galaxies in each subsample and does not alter the velocity dispersion.

 We used $u-r=2.2$ to separate blue and red galaxies in Sections \ref{color_result} and \ref{morphology_color}. Since post-starburst (or E+A) galaxies have $u-r\sim 2.2$ (Goto 2003e), it is a concern that the post-starburst galaxies might be classified as blue galaxies in spite of their lack of on-going star formation activity. However, using a similar SDSS data set, we found that post-starburst galaxies are almost non-existent in cluster regions (in at least bright magnitude of low redshift clusters; Goto 2003e,2005a). Even in the field regions, their fractions are very small ($<$0.1\% of all galaxies). In addition, the analysis obtained with the SFR of galaxies in Section \ref{morphology_SFR} should be unaffected by the post-starburst galaxies.   Therefore, it is unlikely that post-starburst galaxies affect the physical interpretation of our results.

\subsection{Physical Implications}\label{physical}

 Considering that the above systematic effects are unlikely to be responsible for our results, the observed result has significant implication for the galaxy evolution models. 
 In Section \ref{absolute_magnitude}, we found evidence that massive galaxies have smaller velocity dispersion than less massive galaxies.
The only way for massive galaxies to reduce velocity in a cluster is through the energy equipartion by dynamical interaction/friction with less massive galaxies. 
 Since the major merging of galaxies (Mamon et al. 1992) is not significant in a cluster with a large velocity dispersion, this interaction may be less vigorous high speed encounter (galaxy harassment; Moore et al. 1996,1998). 
 In either case, the observed decrease in velocity indicates that  massive galaxies lose their velocity due to the energy equipartion through dynamical interaction/friction.

 Also, the observed result has significant implication for the galaxy evolution models based on the pressure induced gas stripping. The ram-pressure by the cluster ICM is proportional to  the ICM density and square of the galaxy velocity ($\propto \sigma v^{2}$). Therefore, the ram-pressure should be stronger for galaxies with high velocity. Naturally, galaxies with higher velocity are expected to be more stripped and evolved if the ram-pressure stripping (or the hot gas stripping) is responsible for the cluster galaxy evolution. 
 However, the observed trend is opposite; in our analysis, star-forming late-type galaxies show larger velocity dispersion than passive late-type galaxies. Thus, the observed trend suggests that the ram-pressure stripping is not likely to be a dominant physical process responsible for the colour evolution of the cluster galaxies. 

 One might argue that the star-forming late-type galaxies may be still in the process of accretion from the field, and thus, they might have the large velocity dispersion, but have not yet felt the ram-pressure. To avoid the contamination from the still infalling galaxies, we have limited our analysis to those galaxies within 3 $\sigma$ of the velocity dispersion of the cluster they belong to.  
 In addition, the timescale of the ram-pressure stripping is very short ($\sim$50 Myr; Quilis et al. 2000) compared with that of galaxy infall ($\sim$1 Gyr). Therefore, the galaxy should feel the ram-pressure almost immediately it enters the cluster. In this work, the SFR is measured from H$\alpha$ emission line, which should reflect the disappearance of the star-formation activity almost instantaneously ($\sim$50 Myr).  Furthermore, galaxies which have been just accreted should have a lower velocity, i.e., galaxies increase their velocity as they approach the cluster core, falling into the gravitational potential. Therefore, infalling galaxies cannot increase the velocity dispersion of star-forming late-type galaxies.  
 Also, it has been reported that majority of galaxies around the virial radius have already been through the cluster core based on the computer simulations (Moore et al. 2004; Mamon et al. 2004). 
Therefore, it is an unlikely possibility that statistical number of star-forming galaxies happen to be star-forming since they have been just accreted and yet to feel the stripping. 
In case of the hot gas stripping (starvation), the timescale is longer (a few Gyr; Treu et al. 2003) and star-forming late-type galaxies may be still before transformation. However, the existence of passive late-type galaxies with small velocity dispersion cannot be explained by
 the hot gas stripping, neither. The pressure should be still stronger for galaxies with high velocity even in the case of the hot gas stripping.

 On the other hand, the observed trend is perfectly consistent with the dynamical interaction/friction models (Icke 1985; Lavery \& Henry 1988; Mamon 1992; Bekki 1998). If dynamical interaction between galaxies is responsible for cluster galaxy evolution, such interaction happens more frequently when relative velocity of galaxies are smaller (e.g., Ostriker 1980; Mamon 1992), being consistent with passive late-type galaxies having smaller velocity dispersion than bluer counterparts. 
Also, dynamical friction between galaxies can reduce velocity, distributing well-evolved galaxies (passive early-types) at the center of the gravitational potential as seen in subsection \ref{morpholgy_result}. 

In the literature, supportive evidence can be found. It is reported that spiral galaxies have greater velocity dispersion than ellipticals in nearby clusters (Sodre et al. 1989; Stein 1997; Adami et al. 1998,2000). Emission line galaxies are also found to have larger velocity dispersion in ENACS clusters (Biviano et al. 1997) and in 2dFGRS (Lares et al. 2004). Although these studies did not combine morphological and colour information, from which we have obtained critical evidence against the ram-pressure stripping, all of the above studies are consistent with our findings, and thus, can be considered as supporting our own findings.

 For many years, it has been difficult to specify the responsible physical mechanism for the cluster galaxy evolution. However, we have obtained crucial information on the stripping model by combining colour and morphological information together.
 It is important to extend the analysis presented in this paper to high redshift clusters, where more galaxy evolution are on-going.

\section{Conclusion}\label{conclusion}
 
By taking advantage of the large number of spectra available in the SDSS DR2, we have constructed a composite galaxy cluster with an extremely large number of 14548 member galaxies, providing an opportunity to measure velocity dispersion of various subsample of member galaxies in unprecedented precision. Our findings are following:

(i) We found that massive galaxies (brighter in $M_z$) have smaller velocity dispersion (Fig. \ref{fig:absolute_mag}). This is a direct evidence of massive galaxies losing their velocity by dynamical interaction/friction between galaxies through the energy equipartition.

(ii) We found that star-forming late-type galaxies have larger velocity dispersion than passive late-type galaxies. Since the ram-pressure is proportional to $\sigma v^2$, this is an evidence against the ram-pressure stripping models being responsible for the cluster galaxy evolution. Dynamical interaction/friction between galaxies is more consistent with the observed results. 
  
%


\section*{Acknowledgments}

We thank the anonymous referee for many insightful comments, which have improved the paper significantly.
We are grateful to Dr. Ani Thakar for his friendly help in downloading the publicly available SDSS data.

\clearpage


%

\end{document}